# Revisiting Trust in the Era of Generative AI: Factorial Structure and Latent Profiles

Haocan Sun [a,b], Weizi Liu [c], Di Wu [a], Guoming Yu [a]*, Mike Yao [b]*

[a] *School of Journalism and Communication, Beijing Normal University, Beijing, China, 100875*
[b] *Institute of Communications Research, College of Media, University of Illinois Urbana-Champaign, Urbana, IL, USA, 61801*
[c] *Bob Schieffer College of Communication, Texas Christian University, Fort Worth, TX, USA, 76129*
*Corresponding Author:*

**Abstract**

Trust is one of the most important factors shaping whether and how people adopt and rely on artificial intelligence (AI). Yet most existing studies measure trust in terms of functionality, focusing on whether a system is reliable, accurate, or easy to use, while giving less attention to the social and emotional dimensions that are increasingly relevant for today's generative AI (GenAI) systems. These systems do not just process information; they converse, respond, and collaborate with users, blurring the line between tool and partner. In this study, we introduce and validate the Human–AI Trust Scale (HAITS), a new measure designed to capture both the rational and relational aspects of trust in GenAI. Drawing on prior trust theories, qualitative interviews, and two waves of large-scale surveys in China and the United States, we used exploratory (n = 1,546) and confirmatory (n = 1,426) factor analyses to identify four key dimensions of trust: *Affective Trust*, *Competence Trust*, *Benevolence & Integrity*, and *Perceived Risk*. We then applied latent profile analysis to classify users into six distinct trust profiles, revealing meaningful differences in how affective-competence trust and trust-distrust frameworks coexist across individuals and cultures. Our findings offer a validated, culturally sensitive tool for measuring trust in GenAI and provide new insight into how trust evolves in human–AI interaction. By integrating instrumental and relational perspectives of trust, this work lays the foundation for more nuanced research and design of trustworthy AI systems.

**Revisiting Trust in the Era of Generative AI: Factorial Structure and Latent Profiles**

Artificial intelligence (AI) has evolved from rule-driven mechanisms to increasingly autonomous, adaptive systems. Recent advances have brought the rise of generative AI (GenAI) systems capable of producing novel text, speech, images, and other content (Dobriban, 2025; Nyaaba et al., 2024). These systems move AI beyond back-end automation toward expressive, communicative roles that directly engage users. As GenAI applications grow more interactive and conversational, their design and user interfaces increasingly position AI as an embodied or communicative entity rather than hidden background processes.

AI has expanded from reactive execution to proactive, agentic systems, ranging from conversational assistants to autonomous agents and multi-agent networks (Russell & Norvig, 2021, pp. 34–61). This trajectory reflects a broader shift in AI functions: from reactive execution toward proactive, context-sensitive, and increasingly agentic forms of intelligence. With evolving agency, AI systems express social cues at the interactional front end, shaping trust not only as an assessment of task performance but also an evaluation of aims, identity, and relational connection (Rosenthal-von der et al., 2023; Sundar, 2020). Human-facing GenAI systems create simulated social connections through iterative interactions that increasingly resemble interpersonal exchanges (Starke et al., 2024).

This blurring of boundaries between tool and social partner makes users' willingness to rely on GenAI a central determinant of adoption and impact. Yet existing frameworks around trust in AI remain largely grounded in rational, performance-based models such as the Technology Acceptance Model (TAM: Davis, 1989), its AI extensions (Choung et al., 2023), and the three-layered "trust in a specific technology" model developed by McKnight et al. (2021). These approaches, while foundational, do not fully capture the hybrid nature of trust in GenAI, where rational evaluation and relational engagement unfold simultaneously. For agentic, conversational GenAI systems where users engage in ongoing, dialogic exchanges and co-construct outcomes, these relational dimensions are central, making it essential to adopt a hybrid trust framework.

This study examines trust within the emerging human–AI synthetic relationships, where AI functions simultaneously as a technological tool and socially agentic partner. To capture this dual nature, we propose a hybrid approach to measuring trust that incorporates both instrumental and relational dimensions. Using two large-scale, cross-national survey waves from China and the United States, we first develop and validate the Human–AI Trust Scale (HAITS) through a rigorous, multi-stage psychometric process. This variable-centered approach involved item generation from existing technology and interpersonal trust measures, qualitative interviews, exploratory factor analysis, and confirmatory factor analysis, resulting in a four-factor model of trust in GenAI: *Affective Trust*, *Competence Trust*, *Benevolence & Integrity*, and *Perceived Risk*. In the second stage, we adopt a person-centered approach, applying latent profile analysis (LPA) to identify distinct trust configurations to reveal how affective-competence, trust-distrust, and trust in specific technology-institutional trust frameworks coexist and differentiate users. Together, our study provides a validated measurement instrument and a typology of trust profiles that illuminate how trust in generative, conversational AI systems is structured, distributed across populations, and related to usage patterns, cultural context, and psychological characteristics.

To ground our investigation, we begin by reviewing key mechanisms through which trust is formed in novel technologies, emphasizing research on trust transfer—how trust in familiar infrastructures, devices, and applications carries over to newer systems. We then examine, more closely, three layers of technological context: infrastructure, devices, and applications, to clarify where trust cues originate and how they shape user expectations. Building on these foundations, we draw on established frameworks and empirical findings to articulate how trust cues, uncertainty, and structural assurances jointly influence trust formation.

## Technology Trust Transfer: From Familiar Technologies to GenAI

*Trust transfer* theory is a well-established framework that explains how trust formed in one context or object can be carried over to another, particularly when the new object is novel or untested. Originally developed in marketing and e-commerce research, the theory posits that users draw on existing trust relationships—whether with brands, institutions, or technologies—as heuristics to guide decisions about new ones. In the context of AI, this means that users' trust in familiar infrastructures (e.g., the internet), devices (e.g., smartphones), or prior applications (e.g., search engines) can "transfer" to newer systems such as GenAI, shaping their willingness to rely on these systems even before direct experience (Renner et al., 2021; Stewart, 2003). Such transfer can occur through both competence- and affective-based trust (Gong et al., 2020). Regardless of the presence of multiple devices with different functions and forms (Okuoka et al., 2022), task similarity enhances the degree of trust transfer (Soh et al., 2019).

Although rarely discussed in technology studies, trust transfer theory shares conceptual ground with several widely adopted frameworks of trust in human-technology interaction. McKnight et al.'s (2011) three-layer model provides a multi-level framework for understanding *where* trust resides, beginning with a general propensity to trust technology, moving through institution-based assurances, and culminating in trust in specific technologies. Trust transfer theory helps explain how trust can move between these layers, offering a process-oriented view of how prior trust experiences shape trust in novel technologies. Complementing this structural perspective, Sundar's MAIN (Modality, Agency, Interactivity, Navigability; 2008) and TIME (Theory of Interactive Media Effects; 2020) models articulate *how* users rely on heuristic cues embedded in technologies and interfaces to form a range of judgments, including perception of credibility, agency, and ultimately trust. Together, these frameworks illustrate that trust transfer is not a single-step phenomenon but a dynamic process in which cues at the interface level can reinforce or reshape trust across different layers. Rather than proposing a formal theoretical integration, we treat trust transfer as the connective process linking these frameworks. By differentiating the sources of trust cues into three levels: infrastructure, devices, and applications, we clarify *where* trust signals originate and *how* they circulate across McKnight's layers and Sundar's cue-based mechanisms, making the dynamics of technological trust more explicit and measurable.

### *Infrastructure*

In information systems research, the concept of infrastructure has evolved from being understood as a technical substrate to a more relational and experiential phenomenon. Early studies described infrastructure as the underlying technical system that enables practices to run on top of it (Star & Ruhleder, 1996). Star and Ruhleder (1996) introduced the relational view, suggesting that infrastructure occurs when technologies and practices are closely entangled and connected. Infrastructure is a transparent socio-technical system. It is embedded in everyday practice, extends beyond a single event, builds on existing bases, and is most visible when it breaks down. Reimers, Schellhammer, and Johnston (2022) extend this view by focusing on individual experience, arguing that infrastructure is best understood as a *home*. In this framework, infrastructure is the relation between a person and their digital world: technologies and practices only become infrastructure when they are absorbed into one's world as familiar equipment and as part of one's identity and purpose.

At the infrastructure level, trust reflects a broad, background sense of whether the entire technological environment can be relied upon. For example, whether users feel confident in the stability of internet networks, cloud platforms, algorithms for large language models, or data security regimes that form the background of their digital life. Because infrastructure underpins all subsequent technology use, trust at this level provides a foundation for trust transfer to devices and applications. Understanding infrastructure as a *home* helps explain why trust on infrastructure is about technical reliability and the sense of familiarity and continuity that allows individuals to transfer their trust in their digital life.

*Devices*

Devices refer to the physical embodiments of technology. According to Nass and Mason (1990), they belong to *technology as a box*, which does not vary with task conditions. Devices emerge only in technologies with a physical entity, such as phones, computers, and robots, which require concrete *form* and *size* (Nass & Mason, 1990).

Device cues shape trust primarily through two kinds of presence, physical presence and social presence. Physical presence refers to tangible, perceptible features of the device, such as its size, form, mobility, texture, or anthropomorphic elements (e.g., eyes, facial features), which contribute to users' perceptions of capability, familiarity, or credibility (Hancock et al., 2011). Social presence, by contrast, refers to cues that evoke interpersonal or emotional resonance beyond mere functionality. In embodied systems such as robots, tactile or haptic contact may act as one form of social presence: for example, a light touch or comforting contact can carry emotional meaning, foster connection, and influence trust judgments (Valori et al., 2024). However, such contact is sensitive: if the touch feels inappropriate or overly invasive, it may reduce trust rather than increase it. In the context of GenAI systems (which may not have physical embodiment), the equivalent of "social touch" might be more metaphorical, for example, simulated gesture like emoji, tone cues, or "contact-like" affordances in user interfaces (Hancock, et al., 2024; Zhang & Lu, et al., 2025), but the underlying logic is the same: cues embedded in the device layer can impact trust.

Even without specific applications, devices can evoke affective bonds through ownership and personal meaning, creating a foundation for trust cues that applications later amplify. For instance, possession and being possessed (static ownership) can evoke nostalgia (Niemeyer & Siebert, 2023). Prolonged use of childhood electronic devices may generate emotional attachment, dependence, and even feelings of desire (Brembeck & Sörum, 2017; Moran, 2002). Devices often carry symbolic meaning, linking personal memories across generations and identities (e.g., *Data Double*) (Baxter, 2016; Stald, 2008).

*Applications*

Applications, which operate atop infrastructure and devices, are where task-specific trust is most directly formed. Unlike infrastructures, applications are visible and do not depend on a fixed device (if the device itself poses no barrier, they can operate across various devices). They are the most direct task actors, and the objects interact with users. Therefore, applications are the broadest source of cues and the only deliberate source of action. Without applications, the infrastructure cannot maintain its invisibility, sustain practices, or be integrated into the home of digital life (Reimers et al., 2022; Star & Ruhleder, 1996). They are the most visible and deliberate layer where trust transfer manifests, as users interact directly with system interfaces and outputs, and create the possibility of task-specific interaction within devices. In interactions with agentic AI systems, trust cues arise from interfaces and algorithms; as users engage, these cues reinforce one another and extend into social exchange and collaboration (Sundar, 2020).

*Transfer trust to GenAI*

Accordingly, GenAI functions as a meta-application: it is built on Large Language Models (LLMs) and broader internet infrastructure, integrating the capabilities of multiple existing applications, and deployed through various devices. Because of this layered nature, trust in GenAI is shaped not only by its own performance but also by the reliability of the infrastructure and devices through which it is delivered (Glikson & Woolley, 2020; Renner et al., 2021; Stewart, 2003).

GenAI aggregates social cues at the interactional front end, using natural language, multimodal interfaces, and APIs that represent the most frequent and salient points of human contact with AI. At the device and infrastructure levels, GenAI retains audiovisual modality and multiple agencies, which can be embodied in robotic entities, and operates across computers and smartphones. Its most significant breakthrough, however, lies in the high interactivity and navigability that enable richer, more iterative exchanges than previous generations of chatbots and social robots (Rosenthal-von der et al., 2023). Earlier AI systems (*Simple Reflex*, *Model-Based*, and *Goal-Based Agents*) were largely task-specific (Russell & Norvig, 2021, pp. 34–61). By contrast, GenAI systems act as *utility-based learning agents*, offer task-based assistance and social, emotional (De Freitas et al., 2025), and creative interactions (Hwang et al., 2024), positioning them as collaborators rather than mere tools (Sundar, 2020). Viewed through the *Computers as Social Actors* (CASA) framework, GenAI systems can take on diverse roles, functioning as an employee, colleague, or assistant (Nyaaba et al., 2024). Indeed, large language model-based agents such as GPT-4 have been shown to simulate human trust behavior, including behavioral factors and action dynamics (Xie et al., 2024). These interactions foster synthetic relationships, quasi-social bonds with AI agents, that can complement or even substitute for human connections (Yankouskaya et al., 2024).

## Dual Dimensions of Trust in GenAI: From Tools to Partners

To further theorize trust in GenAI systems, we distinguish between two complementary dimensions: instrumental trust and relational trust. Instrumental trust emphasizes rational and systematic assessments of a system's capabilities, focusing on perceived reliability, competence, and performance (Lee & See, 2004). Relational trust, in contrast, is rooted in social interactions characterized by emotional vulnerability and mutual responsiveness (Lewicki et al., 1998; Mayer et al., 1995). Instrumental trust is relatively static, representing a snapshot of a user's confidence in a system, but may vary between different tasks, whereas relational trust is dynamic, capable of evolving through iterative dialogues and affective exchange. Empirical studies consistently support this distinction: Competence vs. affective trust (Glikson & Woolley, 2020) and rational vs. affective trust (Stoltz & Lizardo, 2018) map directly onto our proposed dimensions, providing converging evidence that these two trust logics are well recognized across literatures, aligning with the warmth–competence framework in social cognition (Fiske et al., 2007).

Early acceptance studies conceptualized technology adoption, and by extension trust, largely through rational, performance-oriented evaluation, emphasizing perceived usefulness and ease of use as the primary drivers of adoption (TAM; Davis, 1989). More recent AI-focused extensions of this line of research retain this core but add trust as a determinant and differentiate functionality-oriented vs. human-like facets of trust within the acceptance pathway (Choung et al., 2023). In parallel, Sundar's MAIN and TIME frameworks explain how interface features and affordances trigger cognitive heuristics that shape user judgments, including credibility and trust (Sundar, 2008; Sundar, 2020). And consistent with CASA, social cues can elicit mindless anthropomorphism, prompting interpersonal responses to machines, including feelings of social presence and trust (Gambino et al., 2020; Xu et al., 2022).

However, across these bodies of work, trust is still modeled primarily as a mindless response to application-level cues applied to systems presumed to be rule-governed and objective; the "socialness" largely arises from users' heuristic processing rather than from genuinely relational, co-constructed dynamics governing human-to-human social interactions (Kim & Sundar, 2012; Nass & Moon, 2000). GenAI's agentic, conversational exchanges challenge this assumption: repeated dialogue can not only merely trigger, but also develop trust, while proactively employing interpersonal scripts, thereby necessitating a framework (and measure) that captures both instrumental evaluation and relational growth (Starke et al., 2024).

So far, studies have approached human–AI interactions through an interpersonal lens. The empirical work extends interpersonal trust processes to human–AI relationship construction (Sundar & Lee, 2022; Ogawa et al., 2019) and often employs interpersonal trust scales to measure trust in AI (Park & Yoon, 2024). Theoretically, the Human–Computer Trust Model (HCTM) integrates constructs such as perceived risk, reciprocity, benevolence, and competence—elements originally developed for interpersonal trust frameworks (Gulati et al., 2019; Mayer et al., 1995). Kühne & Peter (2023) conceptualize anthropomorphism, different from CASA, as not mindless, but explicit cognition which can be consciously accessed, leading users to attribute personality traits and moral values to technology (Kühne & Peter, 2023). As AI systems produce increasingly rich and interactive cues (Kosinski, 2023), these actively implemented interpersonal scripts will become more consequential, supporting the claim that human–machine communication should be theorized as psychologically and socially equivalent to human–human communication (Banks & de Graaff, 2020, pp. 29-30).

Taken together, theoretical and empirical insights from different lines of work converge on a crucial challenge: how should we conceptualize trust in GenAI systems that are simultaneously evaluated as technologies and as social partners? We maintain that trust in AI is not an either-or proposition. When a user engages with a conversational system like ChatGPT, asking a question, receiving an answer, and probing further, each turn simultaneously elicits instrumental trust (e.g., "Is this answer accurate, coherent, and useful?") and relational trust (e.g., "Does this system seem attentive, reliable, and trustworthy as a partner?"). These judgments are not static snapshots but evolve dynamically with each interaction: a system that delivers accurate answers may strengthen instrumental trust, which in turn can amplify willingness to engage socially, while moments of perceived misunderstanding can erode relational trust even if task performance remains high. This simultaneity and mutual influence underscore the need to adopt a relational viewpoint of trust.

In addition to this theoretical challenge, we also face an empirical one. Trust in AI has typically been measured through real-system interactions, controlled "Wizard of Oz" studies, or hypothetical scenarios. While these approaches have provided valuable insights, they often yield fragmented results and fail to capture the synthetic nature of trust in GenAI systems, where instrumental and relational trust may co-occur and evolve across repeated interactions. Existing instruments frequently focus on functional reliability or single-dimension acceptance metrics, overlooking multidimension and relational dynamics. Moreover, many scales do not account for key affordances of GenAI, such as autonomy and interactivity, nor do they consistently apply rigorous psychometric validation procedures, including adequate sample sizes, independent development and validation samples, and measurement invariance testing (Gillath et al., 2021; Kim et al., 2021; Liu, 2021).

Therefore, these theoretical and empirical gaps highlight the need for an integrated framework and corresponding measurement strategy that captures both the instrumental and relational dimensions of human–AI trust in the GenAI era. As such, we pose the research question:

RQ1: What is the underlying structure of Human–AI Trust in the GenAI era, capturing both instrumental and relational dimensions?

## Trust as a Dynamic Process

While identifying the underlying duality of human–AI trust is critical, it is equally important to recognize that not all users combine these dimensions in the same way. Different people bring different expectations, prior experiences, and levels of skepticism to their interactions with GenAI, leading to distinct patterns of trust and distrust. Rather than assuming population homogeneity, a person-centered approach allows us to group individuals with similar trust profiles, revealing nuanced combinations of multidimensional nature that a purely variable-centered approach might overlook.

Lewicki et al.'s (1998, 2006) framework defines distrust as a confident negative expectation. Therefore, distrust is a conscious, negative evaluation of a technology's intentions, reliability, or ethics has a close bond with perceived risk, emphasizing that trust is a dynamic process, rather than a static pattern (Scharowski et al., 2024). Classic risk research echoes this statement, which shows that people's perception of risk does not remain static but shifts as familiarity and perceived control increase (Slovic, 1996). This insight has been applied to technology trust, where confidence may initially grow with exposure and then plateau or even decline when users encounter complexity or loss of control (William & Adrian, 2006). Therefore, trust and distrust are not simply opposite ends of a continuum: low distrust does not necessarily indicate high trust but may reflect a lack of sufficient evaluative cues (Dimoka, 2010; Lewicki et al., 1998). Lewicki et al. (1998, 2006) further highlight that trust and distrust can coexist, producing psychologically complex states. This is particularly relevant for new, black-box technologies such as GenAI.

Building on McKnight et al.'s (2011) three-layer model, we argue that trust develops through multiple pathways, beginning with general technology propensity and institutional assurances and extending to evaluations of specific systems (Zhang & Lu, et al., 2025). Our scale captures this multi-layered process at the level of user perceptions, but users may weigh these layers differently. By examining distinct combinations of trust dimensions, we can illuminate (a) the possibility of coexistence between trust and distrust and (b) the heterogeneity of trust formation pathways across users. This effort motivates our second research question:

RQ2: What distinct trust profiles emerge in users' perceptions of GenAI?

## Current study

To fully address the theoretical and empirical challenges identified in our review, a comprehensive framework of human–AI trust will ultimately need to integrate instrumental and relational dimensions and capture how they operate across users and contexts over time. As a first step toward that goal, this paper focuses on conceptualization and operationalization, defining trust as comprising instrumental and relational dimensions and empirically examining whether these dimensions form a coherent structure and how they combine across different types of users.

We introduce the Human–AI Trust Scale (HAITS) and validate it through a multi-phase research program. Phase 1 included three sequential studies that together comprise the scale-development process. Study 1 focused on item generation and content validation: we built an initial item pool based on trust transfer theory, existing technology, and interpersonal trust measures, and insights from qualitative interviews to ensure coverage of GenAI's unique relational and competence-based features. Study 2 fielded this item pool in a large-scale cross-national survey in China and the U.S. (Wave 1) and conducted an exploratory factor analysis (EFA) to identify the latent structure. In Study 3, we conducted a second large-

8 Revisiting Human–AI Trust in the Era of Generative AIscale survey with an independent validation sample (Wave 2) and performed confirmatory factor analysis (CFA) to test the stability and generalizability of the factor structure across cultural contexts.

Phase 2 applied a complementary, person-centered lens to the validated Wave 2 data. Using LPA, we identified distinct configurations of trust and distrust that represent how competence, affective trust, and perceived risk combine across users. We then used multinomial logistic regression to examine predictors of profile membership. This phase does not involve new data collection but deepens our understanding by moving beyond variable relationships to reveal heterogeneity across user groups.

Together, these two phases offer a comprehensive approach: Phase 1 addresses the latent structure of trust (variable-centered), while Phase 2 highlights population-level diversity in trust configurations (person-centered), laying a foundation for future research that can examine longitudinal and relational dynamics more directly.

Phase 1: Scale Development and Validation (Variable-Centered)*Method*

*Item generation*

Guided by trust transfer theory, existing technology and interpersonal trust measures, and we incorporated 63 items from the following sources: The Trust Scale for Explainable AI scale (Perrig et al., 2023), the Trust between People and Automation scale (Perrig et al., 2023), the Human–Computer Trust Model (Pinto et al., 2022), Social Service Robot Interaction Trust Scale (SSRIT, Chi et al., 2021), General interpersonal trust (du Plessis et al., 2023) and specific interpersonal trust (Johnson-George & Swap, 1982). As SSRIT originally included both formative and reflective dimensions, the item pool retained only the 3-item reflective Technology Attachment component. To avoid redundancy, we selected the emotional trust component from Johnson-George and Swap (1982). Since GenAI is a typical technology driving the evolution of Human–AI relationships (Starke et al., 2024), we adapted the wording and explained the meaning of GenAI. Redundant items were removed to ensure coherence.

To supplement AI trust scales and identify content potentially overlooked in the GenAI era, we conducted 12 semi-structured In-Depth interviews (age Mean = 26.08, SD = 3.22; eight female) and two focus group discussions (FGDs). Participants were recruited through convenience sampling. The IDI, each lasting 30 to 45 minutes, centered on participants' personal experiences with GenAI and their perceptions of trust in these systems. The two FGDs each involved four participants and lasted approximately 40 minutes. Interview and FGD data were immediately coded after each session by two trained researchers with backgrounds in psychology and communication studies. This iterative coding process was guided by both inductive insights and deductive mappings to existing trust-related constructs (Proudfoot, 2022). We also treated the results of the thematic analysis as exploratory evidence, which was subsequently compared with the statistical findings. Data collection continued until theoretical saturation was reached, when no new themes emerged in subsequent interviews. We found that most newly generated items could be merged with existing items, resulting in nine novel items that could not be integrated. Expert feedback was incorporated to refine and finalize the initial 72-item pool (Carpenter, 2018, see Supplementary Table 4). All English scales underwent a standardized forward and backward translation process (Beaton et al., 2000). Before distribution, interviews and pretests were conducted, and the final questionnaire revealed no ambiguities.

*Sample size*

Following Carpenter's (2018) standards, we created a pool of items three times larger than the anticipated final item count. The ideal participant-to-variable ratio is 20:1. We initially generated 72 items related to human–AI trust, aiming to refine this to a 22-item scale. Therefore, we obtained 1,440 samples and plan to increase this to 1,600 to account for a 10% buffer, equally divided between U.S. and Chinese participants. Since scale development and validation require distinct samples (Carpenter, 2018), the second round focused on the 22-item scale. To assess criterion validity, we incorporated additional items, bringing the total to around 150. With a 10:1 participant-to-variable ratio, we estimated a sample size of 1,500, with 750 per country.

*Participants*

Trust in AI is widely recognized as culturally contingent (Scherr et al., 2025). Cultural dimensions such as individualism-collectivism, uncertainty avoidance, and power distance shape trust dynamics (Chien et al., 2018; Hancock et al., 2011). Therefore, we take China and the U.S. as focal cases, as China and the United States distinct in many cultural dimensions such as individualism-collectivism, power distance, indulgence, and uncertainty avoidance (Hofstede, 2001, pp. 209-278), providing ideal contexts for empirical comparison. Notably, both countries are now also at the forefront of AI adoption and development (Chakravorti et al., 2023).

Table 1. Demographic Description and Cross-National Comparison

| Variables | China (N = 759) | | United States (N = 787) | |
|---|---|---|---|---|
| Exploration Phase | Mean (N) | SD (%) | Mean (N) | SD (%) |
| Age (Years)*** | 31.11 | 7.18 | 37.63 | 12.88 |
| Gender (Female), N (%)*** | 515 | 67.85% | 452 | 57.43% |
| Education Level*** | 5.20 | 0.59 | 4.22 | 1.42 |
| Importance of GenAI in Daily Life*** | 5.58 | 1.05 | 2.94 | 1.25 |
| Validation Phase | China (N = 700) | | United States (N = 726) | |
| Age (Years)*** | 31.70 | 7.53 | 41.71 | 20.06 |
| Gender (Female), N (%)*** | 475 | 67.85% | 374 | 51.51% |
| Education Level** | 5.11 | 0.66 | 4.86 | 1.37 |
| Importance of GenAI in Daily Life*** | 5.77 | 0.78 | 4.87 | 1.78 |
| Usage Frequency*** | 3.98 | 1.48 | 4.74 | 1.75 |
| Workload Proportion (%)*** | 54.42 | 17.34 | 58.37 | 23.50 |
| Use Frequency (Usefre4)*** | 13.91 | 2.58 | 12.44 | 4.23 |
| Outgroup Trust*** | 7.86 | 1.92 | 9.14 | 4.34 |
| Technical Proficiency*** | 26.80 | 4.63 | 25.66 | 6.12 |
| Technology Readiness** | 78.86 | 8.55 | 77.12 | 12.40 |
| Technology Acceptance | 52.58 | 3.11 | 52.17 | 6.79 |
| Literacy*** | 110.57 | 14.38 | 109.20 | 18.96 |

Note. $p < .05$. $p < .01$. $p < .001$. Significance levels indicate results of *t*-tests comparing Chinese and U.S. samples on each variable.

Chinese participants were recruited through Credamo, an established online survey panel provider in China. Eligible participants were Chinese residents aged 18 years or older. U.S. participants were recruited through Prolific. Eligible participants were U.S. residents aged 18 years or older. Participants completed a self-administered questionnaire between November 2024 and April 2025. In the exploratory phase, the total sample included 1,546 respondents, of whom 64 were excluded. The Chinese sample consisted of 759 adults (Mage = 31.11, SD = 7.18; women = 67.85%), while the American sample included 787 adults (Mage = 37.63, SD = 12.88; women = 57.43%). In the validation phase (n = 1,426), an additional 35 participants were excluded due to failure in attention checks or manipulation checks (e.g., "I have never used a GenAI"). The Chinese sample consisted of 700 adults (Mage = 31.70, SD = 7.53; 67.85% women), and the American sample included 726 adults (Mage = 41.71, SD = 20.06; 51.51% women).

*Measurement*

The measurements consist of three components: Technology Use & Attitudes, Individual Traits & Symptoms, and Control Variables (Table 2).

Table 2. Measure description of variables

| Variables | Measure Description | Source and Scale | Cronbach's alpha |
|---|---|---|---|
| Technology Use & Attitude | | | |
| Daily GenAI Use Frequency | Single-item measure; respondents indicated how often they used the 6 most popular GenAI tools daily. | Adapted from Van Den Eijnden et al. (2016); 8-point frequency scale: 0 (never) to 7 (>40 times/day). | NA |
| Addiction | Single-item measure: "To what extent do you feel addicted to ?" | Self-developed; 5-point scale: 1 (not at all addicted) to 5 (strongly addicted). | NA |
| Literacy | 28-item scale measuring five dimensions: Technical Proficiency, Critical Evaluation, Communication Proficiency, Creative Application, and Ethical Competence. | Liu et al. (2025); 5-point Likert scale: 1 (not at all) to 5 (very much). | 0.946 |
| Technology Acceptance | 12 items assessing perceived usability and ease of use (e.g., "Easy to manipulate"). | Chin et al. (2008); 8-point bipolar scale. | 0.85 |
| Technology Readiness | 18 items capturing innovativeness, optimism, discomfort, and insecurity. | Lam et al. (2008); 5-point Likert scale: 1 (strongly disagree) to 5 (strongly agree). | 0.819 |
| Individual Traits & Symptoms | | | |
| Self-Construal | 12 items measuring independent and interdependent self-construal. | Gudykunst & Lee (2003); 5-point Likert scale: 1 (strongly disagree) to 5 (strongly agree). | 0.889 |
| Outgroup Trust | 2 items: "Most people can be trusted" and "Trust strangers without direct pecuniary interest." | Feng et al. (2016); 7-point Likert scale. | NA |
| Attention Deficit | 6 items from DSM-IV ADHD checklist (e.g., "I am easily distracted"). | Kessler et al. (2005); 5-point Likert scale: 1 (never) to 5 (very often). | 0.864 |
| Loneliness | 8 items from UCLA-8 (e.g., "I feel left out"). | Hays & DiMatteo (1987); 4-point Likert scale: 0 (never) to 4 (always). | 0.861 |
| Perceived Stress | 4 items (e.g., "Unable to control important things in life"). | Cohen et al. (1983); 5-point Likert scale: 0 (never) to 4 (very often). | 0.763 |
| Self-Esteem | 10-item short version of the Rosenberg Self-Esteem Scale. | Rosenberg (1965) 5-point Likert scale: 1 (strongly disagree) to 5 (strongly agree). | 0.864 |
| Control Variables | | | |
| Nationality | Country of citizenship. | Coded as: China=1 / U.S.=0 | NA |
| Age | Calculated as: 2025 - year of birth. | Self-reported birth year. | NA |
| Gender | Participants' identified gender. | Female (1), Male (2), Others (3). | NA |
| Education | Highest level of education completed. | 1 = Less than high school, …, 8 = Doctoral degree. | NA |

Note: NA = Not applicable.

*Analysis strategy*

Considering the discussion on omega and alpha (Hayes & Coutts, 2020; Raykov et al., 2024), we simultaneously computed both alpha and omega (composite reliability) as measures of scale reliability and as references for convergent validity. The extraction method used in EFA was Principal Axis Factoring; we used geominQ and oblimin rotations to cross-validate and allow factor correlations and to ensure robustness under large samples with potential noise (Nguyen & Waller, 2023). The number of factors retained was determined based on the results of parallel analysis, absence of cross-loadings or low-loading items, and theoretical interpretability.

The CFA used Mplus 8.3 to validate the HAITS. Following recommended practices (Sun & Tang, 2025), we used Maximum Likelihood parameter estimation with Robust standard errors (MLR) to compute factor correlations and assess model fit with the following criteria: Comparative Fit Index (CFI) cutoff of 0.90, Tucker-Lewis Index (TLI) cutoff of 0.90, Root Mean Square Error of Approximation (RMSEA) cutoff of 0.07 (Steiger, 2007), and Standardized Root Mean Square Residual (SRMR) cutoff of 0.08 (Hu & Bentler, 1999). Discriminant validity was assessed using the Heterotrait-Monotrait Ratio of Correlations (HTMT, Henseler et al., 2015), with values and bootstrap-based lower 90% and upper 95% confidence intervals < .80 considered acceptable (Cheung et al., 2023; Voorhees et al., 2016). Convergent validity standards include standardized factor loadings not less than 0.5, and composite reliability (Omega) greater than 0.7 (Cheung et al., 2023). Following established procedures (Putnick & Bornstein, 2016; Vandenberg & Lance, 2000), we evaluated configural, metric, scalar, and residual invariance across nationality and gender. Model comparisons followed standard $\Delta$AFI criteria: $\Delta$CFI $\geq$ -.010 and $\Delta$RMSEA $\leq$ .015 indicate support for invariance (Putnick & Bornstein, 2016).

We also conducted a sensitivity analysis using EGA to verify the number of dimensions (Details are provided in the Supplementary), since EGA requires no rotation and more accurately identifies the dimensions (Golino et al., 2020).

*EFA Results*

Parallel analysis supported four- and five-factor solutions (see supplementary Factor Retention). GeominQ and oblimin rotations revealed similar results. Among them, the four-factor model demonstrated the best balance between statistical improvement, structural simplicity, and theoretical interpretability. We systematically eliminated items with cross-loadings or low factor loadings. With an item pool of 54, most items loaded strongly on a single factor with no substantial cross-loadings, yielding a clear and interpretable factor structure.

To guarantee a one-third ratio of the final scale to the initial item pool, we conducted an additional round of screening to remove semantically redundant items. This refinement process led to a final set of 22 items (Factor, Pattern, and Structure Matrix details see Supplementary Table 6 and 7). The four-factor structure remained intact throughout the process. The final solution (see Table 3) explained 56.74% of the total variance.

Table 3. Description of HAITS items and results of EFA and CFA

| Dimensions and Items | EFA FL | Extraction | CFA FL | SMC | Omega | AVE |
|---|---|---|---|---|---|---|
| **Affective Trust** | | 41.462 | | | 0.943 | 0.735 |
| I feel that the AI is a part of me | 0.971 | 0.759 | 0.879 | 0.773 | | |
| AI is my close friend | 0.971 | 0.794 | 0.912 | 0.832 | | |
| There is a lot of warmth in the relationships between the AI and me. | 0.564 | 0.414 | 0.894 | 0.799 | | |
| I like using AI for chatting to relax | 0.800 | 0.619 | 0.809 | 0.654 | | |
| I think that AI is competent and effective in providing emotional support | 0.761 | 0.678 | 0.808 | 0.653 | | |
| My needs and desires are very important to AI | 0.702 | 0.622 | 0.836 | 0.699 | | |
| **Perceived Risk** | | 8.453 | | | 0.959 | 0.683 |
| The AI is deceptive | 0.760 | 0.572 | 0.827 | 0.684 | | |
| The AI behaves in an underhanded manner | 0.729 | 0.442 | 0.737 | 0.543 | | |
| I am suspicious of the AI's intent, action, or, outputs | 0.756 | 0.594 | 0.798 | 0.637 | | |
| The AI's actions will have a harmful or injurious outcome | 0.745 | 0.566 | 0.804 | 0.646 | | |
| It is risky to interact with AI | 0.582 | 0.443 | 0.772 | 0.596 | | |
| **Competence Trust** | | 4.065 | | | 0.966 | 0.626 |
| The AI is reliable | 0.665 | 0.603 | 0.787 | 0.619 | | |
| I think that AI is competent and effective in its performance | 0.712 | 0.554 | 0.76 | 0.578 | | |
| I think that AI performs its role very well | 0.802 | 0.630 | 0.757 | 0.573 | | |
| I believe that AI has all the functionalities I would expect | 0.729 | 0.443 | 0.628 | 0.394 | | |
| I can trust the information presented to me by AI | 0.539 | 0.636 | 0.804 | 0.646 | | |
| The AI is efficient in that it works very quickly. | 0.558 | 0.345 | 0.568 | 0.323 | | |
| **Benevolence and Integrity** | | 2.763 | | | 0.969 | 0.589 |
| I would be able to confide in AI and know that he/she would not discuss my concerns with others | 0.561 | 0.652 | 0.777 | 0.604 | | |
| I could talk freely to AI and know that AI would want to listen | 0.430 | 0.556 | 0.728 | 0.53 | | |
| If AI didn't think I had handled a certain situation very well, he/she would not criticize me in front of other people | 0.693 | 0.522 | 0.581 | 0.338 | | |
| If I told AI what things I worry about, he/she would not think my concerns were silly | 0.762 | 0.539 | 0.617 | 0.381 | | |
| AI would not knowingly do anything to hurt the organization I am part of | 0.441 | 0.501 | 0.679 | 0.461 | | |

Note: the extraction here used oblimin rotations FL = Factor Loading, SMC = Squared Multiple Correlation; AVE = Average Variance Extracted.

*CFA Results*

The CFA model showed an acceptable to good fit: CFI = .956, TLI = .950, SRMR = .045, and RMSEA = .055. Discriminant validity was supported, with all HTMT values below 0.70 and both the lower 90% and upper 95% confidence intervals below 0.75 (Table 4). Convergent validity was also adequate: omega values surpassed 0.90, and all standardized factor loadings were above 0.50 (see Table 3). EGA Community detection replicates four-dimensional trust (See Supplementary Exploratory Graph Analysis).

Measurement invariance across national and gender groups confirmed the scale's applicability (Details see Supplementary Measurement Invariance). Across nationalities, we found support for configural invariance and metric (loading) invariance, but partial invariance and partial residual invariance model by freely estimating intercepts for DIF-flagged items. Finally, although factor variances did not meet strict invariance criteria, the model satisfied factor covariance invariance, suggesting that the inter-factor relationships remained stable across groups.

In terms of gender, all measurement tests are acceptable; no significant gender differences were observed in latent trust levels. ΔCFI and ΔTLI were all below .005, and ΔRMSEA were smaller than .002, indicating excellent model stability and supporting measurement invariance across gender.

Table 4. Factor correlations of HAITS

|  | Affective Trust | Perceived Risk | Competence Trust | Benevolence and Integrity |
|---|---|---|---|---|
| Affective Trust | 0.942 | 0.257 | 0.564 | 0.587 |
| Perceived Risk | 0.265 [0.211, 0.330] | 0.890 | 0.604 | 0.431 |
| Competence Trust | 0.594 [0.554, 0.642] | 0.600 [0.551, 0.658] | 0.862 | 0.633 |
| Benevolence and Integrity | 0.631 [0.587, 0.684] | 0.456 [0.398, 0.526] | 0.662 [0.617, 0.715] | 0.812 |

Note: Below the diagonal are the ML factor correlations from CFA, the 5000 bootstrap lower bound of the 90% and upper 95% confidence interval. On the diagonal are Cronbach's α coefficients. Above the diagonal are the Heterotrait-monotrait ratio of correlations (HTMT).

The HAITS demonstrated good criterion validity (Figure 1). As expected, human–AI trust was significantly associated with users' reported importance ($r = .59$) and frequency ($r = .44$) of GenAI use, as well as their self-reported addiction tendencies ($r = .38$) and collaborative behaviors ($r = .37$), supporting its predictive utility for actual engagement patterns. Correlations with broader individual difference measures, such as GenAI literacy ($r = .59$), technology readiness ($r = .40$), and technology acceptance ($r = .57$), further underscore the scale's ecological validity in digital environments. Notably, human–AI trust exhibited modest associations with personality-related constructs, including interdependence ($r = .38$), self-esteem ($r = .28$), and loneliness ($r = -.31$). Correlations were also found with outgroup trust ($r = .12$), ADHD symptoms ($r = -.09$), and stress ($r = .14$). All correlations reached statistical significance at $p < .001$.

Figure 1. Zero-Order Correlation Matrix Among Key Variables

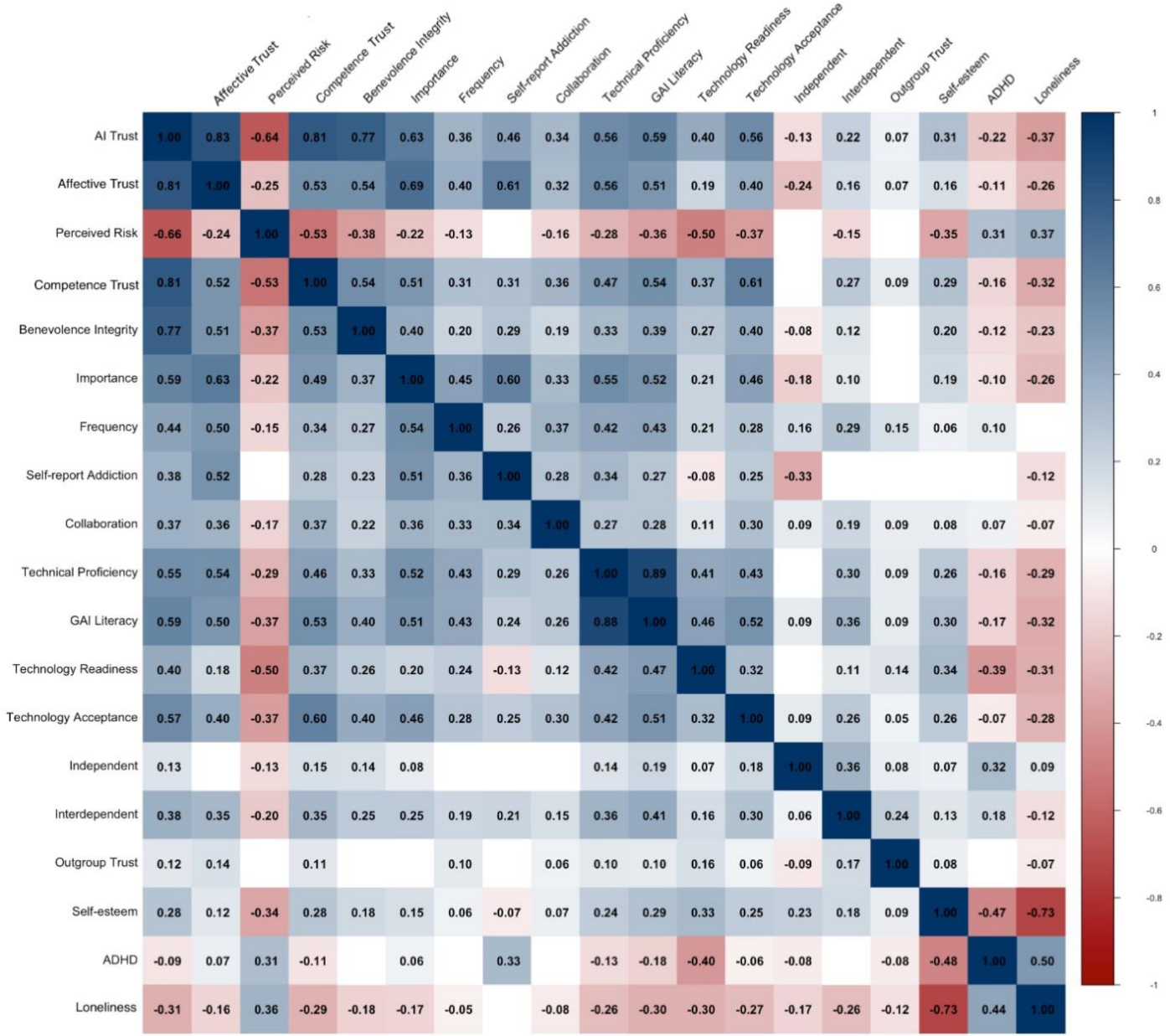

Note. Below the diagonal are partial correlations controlling for nationality, age, and gender. Above the diagonal are zero-order correlations. Blank cells indicate non-significant correlations ($p > .05$).

## Discussion

In our results, we identified four key dimensions: Affective Trust, Competence Trust, Benevolence and Integrity, and Perceived Risk. These dimensions reflect an intersection between the affective-competence trust (Choung et al., 2023), the trust-distrust (Scharowski et al., 2024), and the institutional trust (Mcknight et al., 2011) framework.

As Yankouskaya et al. (2024) observe, GenAI creates pseudosocial bonds that can substitute genuine human relationships. Our statistical results did not reveal a clear representational boundary between interpersonal and Human–AI interactions, suggesting a convergence in how trust is functionally and affectively processed in both contexts. This observation aligns with existing trust measurement trends (Gulati et al., 2019), yet our study further delineates between affective trust and competence trust. The present study is the first to empirically integrate both affective and functional dimensions into the construction of a trust measurement instrument. Our correlation results show that usage frequency and perceived importance of GenAI exert stronger effects on affective trust than on competence trust. This asymmetry extends the dual-process in the HAII-TIME model (Sundar, 2020) to the affective dimension. Compared to competence trust, affective trust in human–AI interaction involves slower and reflective evaluation, which tends to evolve and is shaped by relational depth. *Thick AI* systems (Brandtzaeg et al., 2022) are designed to simulate a personalized companion, facilitating extended interaction, emotional resonance, and even self-disclosure, without the reciprocity in human relationships (Brandtzaeg et al., 2022; Meng & Dai, 2021).

Second, our findings refine the understanding of institutional trust in Human–AI interaction by clarifying the roles of benevolence and integrity within broader trust typologies. Benevolence and integrity reflects the belief that employing a specific class of technologies in each setting is both normal and comfortable. They capture situational normality and structural assurance, core factors of institutional trust, impacting trust in specific technologies, reflected in affective trust and competence trust (McKnight et al., 2011; Lewicki et al., 2006).

Third, the HAITS conceptually and empirically separates trust and distrust as distinct psychological constructs, rather than opposite ends of a continuum. Perceived Risk reflects distrust, a conscious, negative evaluation of a technology's intentions, reliability, or ethics (Scharowski et al., 2024). Supporting this distinction, our findings show that outgroup trust correlates positively with affective and competence trust but not with Perceived Risk, suggesting cognitive separation between trust and distrust (Lewicki et al., 1998, 2006).

## Phase 2: population-level diversity in trust configurations (person-centered)

### Method

### *Analysis strategy*

LPA classifies individuals into distinct groups based on their response patterns, grouping those with similar characteristics. Firstly, we conduct LPA within the validation dataset. Then, for cross-validating the robustness of profiles, we retest the LPA within the development dataset. For LPA, we determined the optimal number of latent profiles by evaluating multiple fit indices, including log-likelihood, Akaike Information Criterion, Bayesian Information Criterion (BIC), sample-size adjusted BIC, and entropy. In addition, the Lo-Mendell-Rubin adjusted likelihood ratio test, the bootstrap likelihood ratio test (5000 times) have been shown to effectively identify the appropriate number of latent profiles (Jenn-Yun et al., 2013). We use multinomial logistic regression to assess predictors across profiles (Liu et al., 2024).

*Result*

Based on both empirical fit indices (see Table 5) and theoretical considerations, we identified the six-class solution as the optimal model to capture the latent heterogeneity within our sample. Class-specific profiles remained consistent across different samples (see Figure 2), with a clear correspondence in key feature patterns, indicating the robustness and replicability of profiles.

Table 5. Latent Profile Model Fit Comparisons in the Exploration and Validation Samples

| Model (Number of Profiles) | Free Parameters | AIC | BIC | aBIC | Entropy | LMR Adj. LRT (p) | BLRT |
|---|---|---|---|---|---|---|---|
| V-3 | 18 | 33115.24 | 33209.96 | 33152.79 | 0.829 | 0.063 | 0.000 |
| V-4 | 23 | 32732.82 | 32853.86 | 32780.79 | 0.839 | 0.003 | 0.000 |
| V-5 | 28 | 32488.30 | 32635.66 | 32546.71 | 0.868 | 0.005 | 0.000 |
| V-6 | 33 | 32310.78 | 32484.45 | 32379.62 | 0.876 | 0.019 | 0.000 |
| V-7 | 38 | 32164.20 | 32364.18 | 32243.47 | 0.832 | 0.057 | 0.000 |
| V-8 | 43 | 32064.24 | 32290.54 | 32153.94 | 0.839 | 0.227 | 0.000 |
| E-4 | 23 | 35750.15 | 35873.05 | 35799.99 | 0.858 | 0.001 | 0.000 |
| E-5 | 28 | 35566.34 | 35715.96 | 35627.01 | 0.839 | 0.054 | 0.000 |
| E-6 | 33 | 35368.70 | 35545.03 | 35440.20 | 0.836 | 0.029 | 0.000 |
| E-7 | 38 | 35266.96 | 35470.01 | 35349.29 | 0.834 | 0.118 | 0.000 |

Note. V denotes the validation phase; E denotes the exploration phase. AIC: Akaike Information Criterion, BIC: Bayesian Information Criterion, aBIC: sample-size adjusted BIC, LMR: Lo-Mendell-Rubin adjusted likelihood ratio test, BLRT: bootstrap likelihood ratio test.

*Moderate Trusters* (24.7%) present a moderate level of competence, affective trust, perceived risk, and benevolence and integrity. As a large and most balanced profile, Moderate Trusters was used as the reference group. In multinomial logistic regression, functional usage frequency, age, gender, and education level did not significantly predict their classification. Emotional usage frequency, importance in their life, technology acceptance, readiness, generative AI literacy, and nationality significantly predicted their classification.

*Full-Spectrum Distrusters* (4.6%) exhibit low trust across affective trust and competence trust, paired with high perceived risk and low benevolence integrity. This group represents low institutional trust, low technology-specific trust, and high distrust. For Class 1, a higher frequency of using GenAI for emotional support predicted lower odds of membership (OR = 0.41, 95% CI [0.26, 0.64], $p < .001$). Greater perceived importance of GenAI (OR = 0.50, 95% CI [0.35, 0.72], $p < .001$) and higher acceptance (OR = 0.84, 95% CI [0.79, 0.90], $p < .001$) also reduced the likelihood of belonging to this class.

*Uncertain Distrusters* (7.4%) report mid-level Competence trust but low institutional and affective trust, suggesting a tentative or emerging orientation toward. This group exemplifies users in the early Stage of trust calibration, reflecting ambivalence rather than outright rejection. Lower odds of membership were associated with emotional-support use (OR = 0.28, 95% CI [0.19, 0.41], $p < .001$). In contrast, higher technology readiness predicted greater odds (OR = 1.04, 95% CI [1.01, 1.07], $p = .003$). U.S. participants were less likely than Chinese participants to be in this class (OR = 5.11, 95% CI [2.33, 11.20], $p < .001$).

*Full-Spectrum Low-Risk Trusters* (53.7%) demonstrate uniformly the highest scores across institutional trust and technology-specific trust. Higher emotional-support use (OR = 1.79, 95% CI [1.51, 2.12], *p* < .001), greater perceived importance of GenAI (OR = 1.30, 95% CI [1.07, 1.57], *p* = .007), higher acceptance (OR = 1.09, 95% CI [1.05, 1.14], *p* < .001), greater technology readiness (OR = 1.06, 95% CI [1.04, 1.08], *p* < .001), and higher literacy (OR = 1.06, 95% CI [1.04, 1.07], *p* < .001) all increased the likelihood of membership. U.S. participants, however, were less likely to belong to this class (OR = 0.49, 95% CI [0.33, 0.72], *p* < .001).

*Rational Trusters* (6.0%) show high confidence in competence and institutional trust but remain affectively distant. With low perceived risk, they constitute trust through a rational cost-benefit appraisal of AI utility. Emotional-support use (OR = 0.56, 95% CI [0.41, 0.76], *p* < .001) and greater importance of GenAI (OR = 0.73, 95% CI [0.56, 0.94], *p* = .015) predicted lower odds of membership, whereas higher acceptance (OR = 1.09, 95% CI [1.02, 1.16], *p* = .007) and greater technology readiness (OR = 1.12, 95% CI [1.08, 1.15], *p* < .001) predicted higher odds. U.S. participants were less likely than Chinese participants to be in this class (OR = 6.88, 95% CI [2.61, 18.17], *p* < .001).

Finally, *Full-Spectrum High-Risk Trusters* (3.7%) express strong affective trust, competence trust, and institutional trust but retain a high level of perceived risk. Emotional-support use (OR = 2.55, 95% CI [1.73, 3.77], *p* < .001) and greater literacy (OR = 1.05, 95% CI [1.02, 1.08], *p* < .001) predicted higher odds of membership. Conversely, higher technology readiness predicted lower odds (OR = 0.92, 95% CI [0.88, 0.96], *p* < .001). U.S. participants were less likely to belong to this class (OR = 3.78, 95% CI [1.33, 10.72], *p* = .013).

Figure 2. Profiles and Distributions Across Four Trust Dimensions

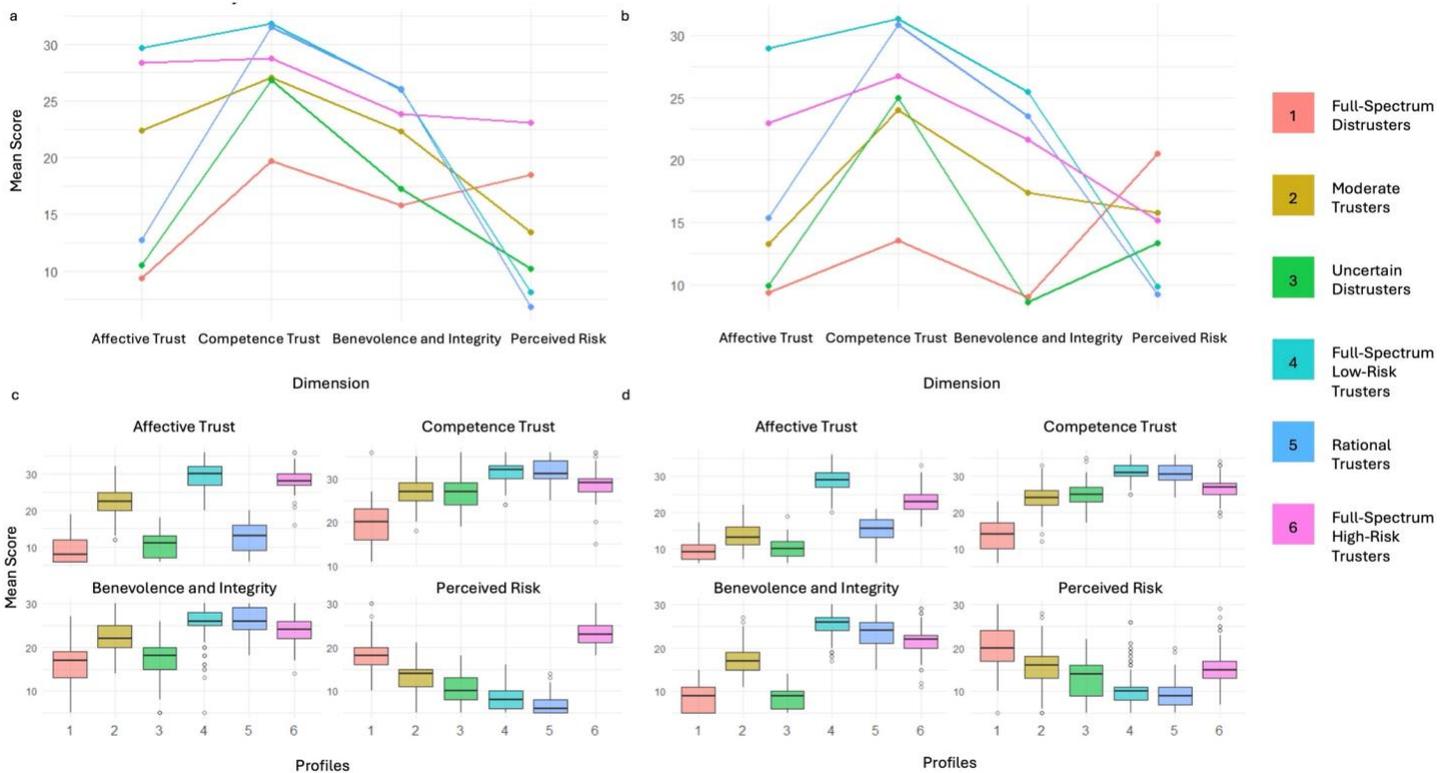

Note. Panels a and c present the results of LPA based on the validation dataset, identifying a 6-class solution. Panels b and d illustrate the characteristics of the six profiles in the exploration dataset. Specifically, a and b are line

plots representing mean scores of the four trust dimensions, while c and d are boxplots visualizing the distribution of each variable across profiles.

## Discussion

Findings from the LPA indicate that competence and affective trust emerge as distinct dimensions. However, competence alone does not adequately predict trust profiles, while affective Trust, reflected in the frequency of using GenAI for emotional support, differentiates even subtle variations across groups. This pattern is consistent with research highlighting the transformative role of anthropomorphized AI in fostering human–AI intimacy (Hu et al., 2025).

Second, comparisons across profiles reveal two consistent patterns. As shown in Rational Trusters, competence trust and affective trust are separate, and high institutional trust does not necessarily translate to high affective trust or competence trust. However, low institutional trust consistently prevents the formation of technology-specific trust, as shown in Systemic Distrusters. This echoes McKnight et al.'s (2011) three-layer cascading framework and can be explained by the fact that low benevolence and integrity provide few transferable positive trust cues, thereby burdening the trust transfer process (Renner et al., 2021; Stewart, 2003).

Distinctions between profiles demonstrate that trust and distrust are separate constructs (Lewicki et al., 1998, 2006). High levels of distrust and trust coexist in Full-Spectrum High-Risk Trusters. In emotionally rich companionship scenarios, users often feel socially engaged yet morally wary: trusting the agent's emotional support while simultaneously distrusting its commercial or persuasive motives (Go & Sundar, 2019; Sundar, 2008). In functionally high-stakes settings such as finance or medicine, users tend to develop strong competence trust due to the system's analytical accuracy and efficiency (Choung et al., 2023). However, they may experience heightened perceived risk arising from algorithmic opacity and costly errors (Shin, 2021; von Eschenbach, 2021). While low levels of distrust and trust coexist in Uncertain Distrusters, representing unfamiliarity with AI, thereby causing a high level of uncertainty. A trade-off between trust and distrust happens in Full-Spectrum Low-Risk Trusters. Although the present study is based on cross-sectional data, the LPA findings suggest potential transitions between trust types. For instance, Uncertain Distrusters resembles early-stage uncertain trust as described by Lewicki et al. (1998, 2006). Full-Spectrum Low-Risk Trusters may display a stable trust configuration. However, these patterns may not be fixed but shift across stages or contexts.

Finally, half of the people are Full-Spectrum Low-Risk Trusters. Cross-national differences further highlight the role of cultural and institutional environments in shaping trust distributions. The findings show that American users are less likely than Chinese users to belong to high-trust groups, while Chinese participants are more concentrated in the Full-Spectrum Low-Risk Trusters category. U.S. users, in contrast, appear more frequently in nearly all other profiles except the Full-Spectrum Low-Risk Trusters. This pattern demonstrates that U.S. participants with higher levels of uncertainty avoidance (Wilczek et al., 2025).

## Limitations and future study

This study is based on cross-national cross-sectional data; thus, the transfer across trust dimensions requires further validation. On the one hand, institutional trust predicts specific forms of affective trust and competence trust in technology (McKnight et al., 2011). On the other hand, perceived risk interacts with other trust dimensions at different phases of Human–AI relational development (Lewicki et al., 1998, 2006). Finally, different trust profiles may shift across contexts and evolve with continued AI use, as users accumulate cues and actions. These dynamics offer fertile ground for integrating our model into the HAII-TIME framework, particularly by elaborating the outcomes component.

## Conclusion

This study advances the theoretical and methodological understanding of trust in GenAI by developing and validating a measurement model HAITS, that integrates affective-competence, trust-distrust, and institutional trust frameworks. Drawing on factor analysis, we demonstrated that in the GenAI era, trust in AI is not reducible to a single evaluative continuum but is better conceptualized as a dynamic system of interrelated psychological constructs, including Affective Trust, Competence Trust, Benevolence and Integrity, and Perceived Risk. The HAITS demonstrated measurement invariance across cultural and gender groups.

Latent profile analysis revealed six distinct trust profiles, exploring the potential transformations between different stages of trust: Full-Spectrum Distrusters, Uncertain Distrusters, Moderate Trusters, Full-Spectrum Low-Risk Trusters, Rational Trusters, and Full-Spectrum High-Risk Trusters. These profiles correspond to trust types observed in real-world settings and echo the framework of the HAITS. Specifically, affective and competence trust emerged as separable features of technology-specific trust; benevolence and integrity functioned as institutional predictors of technology-specific trust; and trust and distrust coexisted as distinct dimensions to define uncertainty. Cross-national differences in profile distribution further underscored the cultural sensitivity of AI trust. Together, these findings position HAITS as a psychometrically rigorous instrument and demonstrate how technology trust evolves in the context of GenAI.

20 Revisiting Human–AI Trust in the Era of Generative AI